\documentstyle[preprint,aps,prl,amssymb]{revtex}

\begin{document}

\title{ Vacuum Nodes in QCD at $\theta=\pi$:\\
Exact Results}

\author{\bf {M. Asorey} and F. Falceto,}
\address{Departamento de
F\'{\i}sica Te\'orica. Facultad de Ciencias\\
Universidad de Zaragoza.  50009 Zaragoza. Spain}

\def\CAG{{\cal A/\cal G}} \def\CO{{\cal O}} 
\def\CA{{\cal A}} \def\CC{{\cal C}} \def\CF{{\cal F}} \def\CG{{\cal G}}
\def\CL{{\cal L}} \def\CH{{\cal H}} \def\CI{{\cal I}} \def\CU{{\cal U}}
\def\CB{{\cal B}} \def\CR{{\cal R}} \def\CD{{\cal D}} \def\CT{{\cal T}}
\def\CK{{\cal K}}
\def\e#1{{\rm e}^{^{\textstyle#1}}}
\def\grad#1{\,\nabla\!_{{#1}}\,}
\def\gradgrad#1#2{\,\nabla\!_{{#1}}\nabla\!_{{#2}}\,}
\def\ph{\varphi}
\def\psibar{\overline\psi}
\def\om#1#2{\omega^{#1}{}_{#2}}
\def\vev#1{\langle #1 \rangle}
\def\lform{\hbox{$\sqcup$}\llap{\hbox{$\sqcap$}}}
\def\darr#1{\raise1.5ex\hbox{$\leftrightarrow$}\mkern-16.5mu #1}
\def\lie{\hbox{\it\$}} 
\def\ha{{1\over2}}
\def\half{{\textstyle{1\over2}}} 
\def\roughly#1{\raise.3ex\hbox{$#1$\kern-.75em\lower1ex\hbox{$\sim$}}}
\def\inbar{\,\vrule height1.5ex width.4pt depth0pt}
\def\IB{\relax{\rm I\kern-.18em B}}
\def\IC{\relax\hbox{$\inbar\kern-.3em{\rm C}$}}
\def\ID{\relax{\rm I\kern-.18em D}}
\def\IE{\relax{\rm I\kern-.18em E}}
\def\IF{\relax{\rm I\kern-.18em F}}
\def\IG{\relax\hbox{$\inbar\kern-.3em{\rm G}$}}
\def\IH{{\Bbb H}}
\def\II{\relax{\rm I\kern-.18em I}}
\def\IK{\relax{\rm I\kern-.18em K}}
\def\IL{\relax{\rm I\kern-.18em L}}
\def\IM{\relax{\rm I\kern-.18em M}}
\def\IN{\relax{\rm I\kern-.18em N}}
\def\IO{\relax\hbox{$\inbar\kern-.3em{\rm O}$}}
\def\IP{\relax{\rm I\kern-.18em P}}
\def\IQ{\relax\hbox{$\inbar\kern-.3em{\rm Q}$}}
\def\IR{{\Bbb R}}
\font\cmss=cmss10 \font\cmsss=cmss10 at 10truept
\def\IZ{\relax\ifmmode\mathchoice
{\hbox{\cmss Z\kern-.4em Z}}{\hbox{\cmss Z\kern-.4em Z}}
{\lower.9pt\hbox{\cmsss Z\kern-.36em Z}}
{\lower1.2pt\hbox{\cmsss Z\kern-.36em Z}}\else{\cmss Z\kern-.4em Z}\fi}
\def\ZZ{{\Bbb Z}}
\def\IGa{\relax\hbox{${\rm I}\kern-.18em\Gamma$}}
\def\IPi{\relax\hbox{${\rm I}\kern-.18em\Pi$}}
\def\ITh{\relax\hbox{$\inbar\kern-.3em\Theta$}}
\def\IOm{\relax\hbox{$\inbar\kern-3.00pt\Omega$}}
\def\CA{{\cal A}}\def\CCA{$\CA$}
\def\CC{{\cal C}}\def\CCC{$\CC$}
\def\CT{{\cal T}}\def\CCT{$\CT$}
\def\CQ{{\cal Q}}
\def\CS{{\cal S}}
\def\CP{{\cal P}}\def\CCP{$\cal P$}
\def\CO{{\cal O}}\def\CCO{$\CO$}
\def\CM{{\cal M}}\def\CCM{$\CM$}
\def\CMH{\widehat\CM}\def\CCMH{$\CMH$}
\def\CMB{\overline\CM}\def\CCMB{$\CMB$}
\def\CH{{\cal H}}\def\CCH{$\CH$}
\def\CL{{\cal L}}\def\CCL{$\CL$}
\def\CS{{\cal S}}\def\CCS{$\CS$}
\def\CX{{\cal X}}
\def\CE{{\cal E}}\def\CCE{$\CE$}
\def\CV{{\cal V}}\def\CCV{$\CV$}
\def\CU{{\cal U}}\def\CCU{$\CU$}
\def\CF{{\cal F}}\def\CCF{$\CF$}
\def\CG{{\cal G}}\def\CCG{$\CG$}
\def\CN{{\cal N}}\def\CCN{$\CN$}
\def\CD{{\cal D}}\def\CCD{$\CD$}
\def\CZ{{\cal Z}}\def\CCZ{$\CZ$}
\def\cs{S_{\rm CS}(A)}
\def\css{S_{\rm CS}}
\def\cst{Chern-Simons theory}
\def\csts{Chern-Simons theories}
\def\Tr{{\rm Tr}}
\def\Nabla{{\Bbb D}}
\def\sph{{A_{\rm sph}}}
\def\asph{{\widetilde{A}_{\rm sph}}}
\def\vac{{A_{\rm vac}}}
\def\be{\begin{eqnarray}}
\def\ee{\end{eqnarray}}
\maketitle
\begin{abstract}

	We show that the vacuum functional of 3+1 dimensional
non-abelian gauge theories vanishes for some classical
field configurations $\psi_0(A)=0$  when the coefficient of
the CP violating $\theta$--term  becomes $\theta=\pi$ (mod. $2\pi$).
Some of these classical configurations are explicitly identified
and  include   sphalerons. The results shed new light
into the non-perturbative behavior of non-abelian gauge theories
and suggest a relevant role for these classical configurations
in the confinement mechanism at $\theta=0$.
\end{abstract}
\hyphenation{}
\pacs{PACS: 12.38.Aw, 11.15.-q, 11.30.Er
{\tt$\backslash$\string pacs\{\}} should always be input,
even if empty.}

 \narrowtext\noindent

In spite of the great phenomenological success of QCD we are
still lacking a good understanding of the non-perturbative behavior
of the theory in the strong coupling regime.
The only known rigorous results on the low energy regime of QCD
concern the existence of fermion mass
inequalities \cite{weig} and the absence of parity symmetry
breaking in vector-like theories
\cite{witvaf}.
However, after so many years there has been little progress in the
analytic understanding of the mechanisms of confinement and chiral
symmetry breaking. In particular, the role of relevant classical
field configurations in those phenomena is not  clear yet.

Very recently a new analytic approach made possible the understanding
of such mechanisms in supersymmetric gauge theories \cite{seiw}.
The results support the t' Hooft-Mandelstam  dual
superconducting scenario for the low energy dynamics of the theory.

In this note, following a different approach we obtain
non-perturbative information on the structure of QCD in the strong
coupling regime. The main idea is to analyse the response of    QCD
vacuum to the effect of  the
$\theta$--term. The same idea has been very  fruitful in the
understanding the vacuum structure of 2+1 dimensional gauge theories
\cite{nod}. In such a case
the effect  of a Chern-Simons perturbation leads to the
appearance of vacuum nodes and deconfined quarks. Some of those
nodes  appear for abelian magnetic monopole configurations which
suggested a relevant role for those configurations in the confinement
mechanisms and supported the  dual
superconducting scenario for quark confinement.
In this note we show that similar results hold for the 3+1
dimensional case.

In the temporal gauge ($A_0=0$) and
Schr\"odinger representation \cite{Gold-Jack}, the quantum
Hamiltonian of
Yang-Mills theory
with a $\theta$--term  is given by
\widetext
\be
 \IH_\theta =\int d^3 x\  \Tr \left\{ {  g^2}\left( {\delta\over
\delta A_k} - {i\theta\over 16
\pi^2}\epsilon_{kjl}F^{jl}\right)
 \right.
\left.  \left({\delta\over \delta A^k}-{i\theta\over 16
\pi^2}\epsilon^{kjl}F_{jl}\right)
 - {1\over 2 g^2}  F^{jk} F_{jk} \right\}.\label{hamq}
\ee
 \narrowtext\noindent
\noindent
 Because of Gauss law constraint
\be
{ D^k_A {\delta\over \delta A_k}\psi_{\rm ph}(A)=0,}\label{gaussss}
\ee
the physical states must be gauge invariant functionals, i.e.
$\psi_{\rm ph}(A^{\phi})=\psi_{\rm ph}(A)$
for any gauge transformation $\phi$
with $A_\mu^\phi=\phi^{-1}A_\mu\phi+
\phi^{-1}\partial_\mu \phi$. Since the physical states are constant
along the gauge orbits, they can be identified with
functionals on the space of gauge orbits $\CM=\CA/\CG$, i.e.
 the quotient space of the space of 3-dimensional
gauge fields $\CA$ by the group of gauge transformations $\CG$.

The same constraint analysis led to the existence of nodes in all
physical states of $2+1$ dimensional gauge theories
with a Chern-Simons term,
 because these states are sections of a non-trivial
line bundle over $\CM$ \cite{kar}. In the present
case the Gauss law is not so stringent  and the existence
of nodes cannot be derived from pure kinematical properties.

Moreover, there is a fundamental technical difference between the two
cases. In 2+1 dimensions the only ultraviolet
divergence appears into the vacuum energy.
Therefore a trivial constant subtraction is enough to renormalize
the quantum Hamiltonian and to give  sense to the Schr\"odinger
formalism. However, in 3+1 the structure of divergences becomes more
involved and one has to renormalize also the wave fields and the
coupling constant. In order to have a control of the ultraviolet
divergences and the renormalization of the theory  we introduce an
ultraviolet  regularization of the Hamiltonian (\ref{hamq})
%
$$
\IH_\theta^{\rm reg}=
\int d^3 x\  \Tr \left\{ {  g^2}\left( {\delta\over \delta A_k} -
{i\theta\over 16
\pi^2}\epsilon_{kjl}F^{jl}\right)(I+{\Delta_A\over\Lambda^2})^{-n}
 \right.\left.  \left({\delta\over \delta A^k}-{i\theta\over 16
\pi^2}\epsilon^{kjl}F_{jl}\right) 
 -{1\over 2 g^2}  F^{jk} F_{jk} \right\}
$$
%
of the type considered in \cite{amm}. In this framework the
Schr\"odinger formalism of the
the quantum theory \cite{amm}\cite {Sym}
becomes as safe   as in the 2+1 dimensional case
\cite{Feynman}.

The non-trivial effect of the $\theta$--term is due to the
non-simply connected character of the orbit space $\pi_1(\CM)=\ZZ$
or what is equivalent the non-connected character of the group of
gauge transformations $\pi_0(\CG)=\ZZ$. In physical terms, for
instance,  monopoles are turned into  dyons by this effect
\cite{Witten}. The regularized
Hamiltonian  can be written as
$$
{ \IH^{\rm reg}_\theta= \int d^3 x\ \Tr \left\{
{g^2}\Nabla^\theta_j (I+{\Delta_A\over\Lambda^2})^{-n}\Nabla_\theta^j
 - {1\over 2 g^2}   F^{jk} F_{jk}\right\},}
$$
  \narrowtext\noindent
where
\be
{\Nabla^\theta_j={\delta\over \delta A_j}-i{\theta\over 16\pi^2}
\epsilon_{jkl}F^{kl},}\label{der}
\ee
is a functional covariant derivative with respect to the U(1) ultra-gauge
field defined over the space of gauge fields $\CA$ by
the ultra-gauge vectior potential
\be
{\alpha^\theta_j={\theta\over
16\pi^2}\epsilon_{jkl}F^{kl}.}\label{form} \ee
The projection of this vector potential into the orbit space $\CM$  is
non-trivial for any value of $\theta\neq 2n\pi$. Actually,
the projection of the potential
$(1/\theta)\alpha^\theta$ is a generating form of the non-trivial
first cohomology group $H^1(\CM,\ZZ)=\ZZ$ of $\CM$. In this sense
the phenomenon is very similar to the Aharanov-Bohm effect, and the
effect of the $\theta$--term cannot be globally removed by a
gauge transformation over the space of gauge orbits \cite{amm}.

It is, however, possible to find a codimension one submanifold
$\CN\subset \CM$ and    perform
a gauge transformation on its complement $\CM-\CN$
 which removes the $\theta$--dependence
of the Hamiltonian. The price to pay is the appearence
of non-trivial
boundary conditions at
$\CN$. In this sense the transformation which is not globally defined on
$\CM$ trades the $\theta$--dependence
of the Hamiltonian by suitable boundary conditions at $\CN$.

Although the  choice for $\CN$ is not unique, it is convenient for
later purposes to choose
$
{\CN=\{[A]\in\CM; S_{\rm CS}=(2n+1)\pi\},}
$
where
\be
{\cs=
{1\over 4\pi}\int d^3x\ \epsilon^{jkl} \Tr\left(A_j\partial_k A_l+
{2\over3} A_j A_k A_l \right) }\label {hcs}
\ee
is the Chern-Simons functional.
It is trivial to show that the gauge orbits of
the open   subsets of gauge fields
$\CB_n=\{A\in \CA; (2n-1)\pi
<\cs< (2n+1)\pi)\}$
completely fill the space $\CM-\CN$. In a similar
manner, $\CN$ is made up of the orbits of the codimension one submanifolds
${\CK_n=\{ A\in \CA; \cs= (2n+1)\pi \}}
$
 of $\CA$.
Then the transformation
$
{\xi(A)={\rm e}^{-{i\theta\over 2\pi}\cs}  \psi_{\rm ph}(A)}
\label{trans}
$
is uniquely defined on the domain $\CB_0$ and  the
quantum Hamiltonian becomes $\theta$--independent
under such a transformation,
$
{\widetilde{\IH}_\theta^{\rm reg}=
{\rm e}^{-{i\theta\over 2\pi}\cs}
\IH_\theta^{\rm reg} {\rm e}^{{i\theta\over 2\pi}\cs}=\IH_0^{\rm
reg}}.$
 The $\theta$
dependence is encoded in the non-trivial boundary conditions that
physical states have to verify now at the boundaries $\CK_{-1}$
and $\CK_{0}$  of the domain $\CB_0$,
\be
{\xi(A_+)={\rm e}^{-{i\theta}} \xi(A_-)}\label{bc}
\ee
 for any pair of gauge fields
$A_-\in \CK_{-1}$ and $A_+\in
\CK_{0}$
 which are gauge equivalent
 $ A_+=A_-^\phi$  by a gauge transformation with
winding number $\nu(\phi)=1$.
In this sense the transformation  is trading the
$\theta$--dependence of the Hamiltonian by non-trivial boundary
conditions
on $\CN$.

 We can now identify some physically relevant gauge
field configurations dwelling on this boundary  $\CN$.
In the center of our domain $\CB_0$ we have the classical
Yang-Mills vacuum,
$A_{\rm vac}=0$.
The other n-vacua classical configurations $ A_{\rm n\cdot vac}=
\phi_n^{-1} d \phi_n$ belong
to the different disconnected open domains $\CB_n$ of $\CA$ with $
\nu(\phi_n)=n. $
In the boundary of the domain $\CB_0$ we have
 sphalerons $\sph$ and anti-sphalerons $\asph$.
 They are defined for finite volumes as
static, unstable solutions of Yang-Mills equations with
only one unstable direction. They appear for
any compact 3 dimensional space  $\Sigma$ due to the non-simply
connected character of the gauge orbit space $\CM$ \cite{man}.
 The explicit expression, however is only known for
some special manifolds. In the case of the sphere $\Sigma=S^3$,
the first sphaleron is given by the 3-dimensional gauge
field configuration induced by the 4-dimensional BPST instanton
on the 3-dimensional sphere centered at the center of the instanton and
with radius equal to the size of the
instanton \cite{pvB}.
The unstable mode   corresponds to the radial
perturbation generated by the flow
of the instanton. In stereographic projection coordinates, it has
the following expression
\widetext
\be
A_{\rm sph}{\ }^a_j={4\rho\over
(x^2+(2\rho)^2)^2}\left(4\rho\epsilon^{ajk}
x^k-2 x^a x_j\right.
+  \left.
(x^2-(2\rho)^2)\delta^a_j\right). \label{sp}
\ee
 \narrowtext\noindent
where $\rho$ is the radius of the sphere and also the size of the
instanton.
A relevant property of this sphalerons is that for them
 the Chern-Simons action
$
S_{\rm cs}(\sph)=\pi
$
reachs the value $\pi$. This property which turns out to be crucial for
our analysis
is independent of the size of the sphere because the Chern-Simons
functional is metric independent,
and  implies that $\sph\in \CK_0$. The only change  when we
modify the metric of the sphere, e.g. we increase the value
of its radius $\rho$ to get an infinite volume $\IR^3$, is that the
 configuration
$\sph$ will not be anymore a solution of Yang-Mills equation but
this fact will have no relevance for our discussion. There is another
configuration $\asph$  with similar properties generated by  the
anti-instanton, %
\widetext
\be
{\asph}{\ }^a_j={4\rho\over
(x^2+4\rho^2)^2}\left[4\rho\epsilon^{ajk}
x^k+2 x^a x_j \right.
- \left.(x^2-(2\rho)^2)\delta^a_j\right]. \label{asp}
\ee
 \narrowtext\noindent
 with ${S_{\rm cs}(\asph)=-\pi}$, i.e. $\asph\in \CK_{-1}$.

An interesting connection between these two configurations is
that they are gauge equivalent $\sph=\widetilde{A}^\Phi_{\rm sph}$
under a gauge transformation
\be
{\Phi(x)= {1\over x^2 + 4\rho^2}\left[  (x^2 - 4\rho^2)I - 8
\rho  \tau_j x^j\right].}
\label{gt}
\ee
with winding number $\nu(\Phi)=1$. Therefore, our physical states
satisfy that
${\xi(\sph)={\rm e}^{-{i\theta}}
\xi(\asph)}$
on sphalerons configurations.

Now, the relevant property which will allow us to extract some
information on the quantum vacua nodal structure is that the  theory
is CP invariant for $\theta=\pi$. The theory is C invariant for any
value of $\theta$, but parity is not preserved unless $\theta=\pi$.
In the Hamiltonian approach this property of the special case
$\theta=\pi$ comes from the fact that the boundary condition
(\ref{bc}), becomes an anti-periodic boundary condition,
${\xi(A_+)=- \xi(A_-)}$,
 which is a CP invariant condition, because the gauge fields of
the boundary $\CK_{0}$ are transformed into fields of $\CK_{-1}$
and conversely. This is dictated by
the odd transformation of the Chern-Simons action under parity
(charge conjugation leaves invariant the Chern-Simons  and
Yang-Mills terms).
In fact, we have that  P$\CK_{n-1}=\CK_{-n}$, which also implies that the
submanifold $\CN$ is invariant under parity.

Let us now consider the full Hilbert space of physical states $\CH$
defined exclusively by the anti-periodic boundary condition.
This means that we are not presupposing anything on the possible
existence of superselection sectors in $\CH$ and
spontaneous breaking of CP symmetry. In such a Hilbert space it
is always possible to find a complete basis of stationary wave functionals
with definite CP symmetry. If the energy level is not degenerate
the corresponding physical state $\xi(A)$
has to be CP even or CP odd. In the degenerate case, if $U(P)\xi(A)$
is not on the same ray that $\xi(A)$, the
stationary functionals $\xi_\pm(A)=
\xi(A)\pm U(P)\xi(A)$ are parity even/odd, respectively. Notice that
because of the CP invariance of the anti-periodic boundary
conditions, $\xi_\pm(A)$  also satisfy the same conditions. If
CP is spontaneously broken the  quantum vacua $\xi_0(A)$ in the different
physical phases will not
have a definite parity ($\xi_0\neq\xi_\pm$) and the  Hilbert space will
split into superselection sectors not connected by local observables.

The behavior of sphalerons and classical vacuum configurations under
parity is rather different. The classical vacuum $A_{\rm vac}=0$ is
P invariant
($A_{\rm vac}^P=A_{\rm vac}$) whereas
the sphaleron it is not, but it is  transformed into a gauge
equivalent configuration, i.e.
$ A_{\rm sph}^P={\asph}=A_{\rm sph}^\Phi$, { with} $\nu(\Phi)=1$.
In fact, the whole  submanifold $\CN \in \CM$
is CP invariant, but  the sphaleron has an additional
peculiarity, it is quasi-invariant under parity transformation.

Now, because of  anti-periodic
boundary conditions, we have that
$
U(P)\xi_0( \sph)= \xi_0(\asph)
 = \xi_0(A^\Phi_{\rm sph }) =-\xi_0(\sph).\label{fpr}
$
If the vacuum is parity even this is
possible only if $\xi_0$
%
%
%
vanishes for sphaleron configurations, $\xi_0(\sph)=0,$.
On the other hand, since
$
U(P)\xi_0( \vac)= \xi_0(A^P_{\rm vac })= \xi_0(\vac),
$
if $\xi_0$ is parity odd, it vanishes for classical vacuum
configurations. Now, in the regularized theory there is no
running of the coupling constant and the potential Yang-Mills term
of the Hamiltonian,
is not suppressed in the infrared. This term, thus, gives a finite positive
contribution to the energy of stationary states. Therefore, the
parity even states which vanish at sphalerons cannot have the
same energy as parity odd states which vanish at classical
vacuum configurations where the potential terms attaints its
minimal value. This feature implies that: i) the quantum vacuum state
$\psi_0(A)$ has to be parity even, ii) it vanishes at $\sph$ and iii)
CP symmetry is not spontaneously broken in the regularized theory.

In the above derivation of the vanishing of the quantum vacuum
functional $\psi_0$ for sphalerons, the only relevant ingredients were
the quasi-invariance of these gauge fields under parity
 and the special value of Chern-Simons functional
$\css(\sph)=\pi$ for them. Therefore the
results can be extended for any gauge field satisfying the same
properties. The space of such gauge field configurations is infinite
dimensional. Infinitesimally, the perturbations $A=\sph+\epsilon\tau
+\CO(\epsilon^2)$ of $\sph$
given by $\tau(x)=-\eta(-x)+\Phi\eta\Phi^\dagger(x)+D_\sph\varphi(x)$
preserve both properties, for any perturbation
of gauge fields $\eta$ and any infinitesimal gauge transformation $\varphi$.
Therefore such perturbations
generate an infinite subspace of nodal configurations for
$\psi_0$ in $\CK_0$. The same configurations are also nodes of any higher
energy stationary states with even parity $\psi_{\rm even}(\sph)=0$.
 The nodal configurations of parity odd excited states contain the
classical  vacuum configurations $\psi_{\rm odd}(A_{\rm vac})=0$.

 Since these results hold for any value of
the ultraviolet regulator $\Lambda$ and the coupling constant $g$,
they also hold in the renormalized theory. In the particular case
of  absence of spontaneous breaking of the symmetry
 the result is even more evident because the phenomenon related to the
infrared properties of the theory which are the same for the
regularized and renormalized theories.

 There is an independent proof of  the same results.
The euclidean time evolution kernel of the regularized theory
from an sphaleron to the classical vacuum is
given by \cite{amm}
\widetext
\be
K^\theta_T(\sph,A_{\rm vac})=<\sph|{\textstyle e}^{-TH_\theta^{\rm reg}}|
A_{\rm vac}>
= \int_{{\phantom{LLLLL}\atop{ A(0)=\vac\atop {A(T)=\sph}}}}
 \!\!\!\!\!\!\!\!\!\!\!\!\!\!\!\!\!
\delta A\ \ \ \ \ {\rm exp}\left\{ -{\int_0^T dt \int d^3 x
\CL^{\rm reg}_\theta}\right\}.\label{ker}
\ee
 \narrowtext\noindent
This kernel is dependent on the choice of the U(1) gauge
over $\CM$, but in the pseudo-periodic gauge (\ref{bc})
it is given by the above path integral prescription.
Parity transformation
involves a change in (\ref{ker}) of the boundary conditions of
gauge field
configurations, $A(T)=\asph$  instead of
$A(T)=\sph$. Therefore, for $\theta=\pi$ we have
 $PK^\pi_T(\sph,A_{\rm
vac})=K^{\pi}_T(\asph,A_{\rm vac})$. Now,
the CP transformation of $\CL^{\rm reg}_\theta$
 is  reduced to the change of the $\theta$--term
contribution, which becomes $-i\theta(\half+n)$ instead of
$i\theta(\half+n)$ for any trajectory  $A(t)$ in the space of gauge
fields $\CA$ whose projection on $\CM$ has winding number $n$.
This implies that for $\theta=\pi$, the kernel $K^\pi_T(\sph,A_{\rm
vac})$ is pure imaginary and thus it is parity odd,
\be
PK^\pi_T(\sph,A_{\rm
vac})=-K^{\pi}_T(\sph,A_{\rm vac}).\label{im}
\ee
On the other hand, as discussed above it is always possible to
find in $\CH$ a complete basis of eigenfunctionals
$\psi_n(A)$ satisfying the anti-periodic boundary conditions
with definite parity for
$\theta=\pi$. In terms of such a basis, the
kernel
\widetext
\be
PK^\pi_T(\sph,A_{\rm vac})=
\sum_n U(P) \psi_n(\sph)^\ast U(P)\psi_n(A_{\rm vac}) {\rm e}^{-E_n
T}=K^\pi_T (\sph,A_{\rm vac}),
\label{kerrr}
\ee
\narrowtext
\noindent
 remains invariant under CP
transformations. This is in disagreement with (\ref{im}) unless
$K^\pi_T(\sph,A_{\rm vac})=0$, which  implies that any
 stationary state  $\psi_n$   either vanishes  at
$A_{\rm vac}$ or at $\sph$. By the parity symmetry argument
discussed above, we know that the first possibility occurs for parity
odd states whereas for parity even eigenstates $\psi(\sph)=0$. The
same argument applies to any pair of field configurations
$A_0$ and $A_\pi$ with $\cs=0$ and $\cs=\pi$
and similar transformation properties under parity.  Therefore,
the vacuum  functional $\psi_0$ vanishes for any
field  configuration $A\in \CK_n$ which is
quasi-invariant   under parity.
 In this sense
the path integral analysis confirms the result obtained by the
Hamiltonian formalism.

 On the other hand numerical simulations indicate
that the theory ceases to confine \cite{desy} at $\theta =\pi$.
Thus, one might
conjecture that these gauge configurations of $\CN$ whose contribution
to the vacuum is completely suppressed in the deconfined phase,
 might play  a role in the confinement  mechanism when $\theta\neq\pi$.
 On the other hand the existence of nodes of the vacuum in the
 sphaleron sector confirms the destructive interference between
 the instanton and anti-instanton    tunneling
 at $\theta=\pi$, first indicated by semiclassical
 instanton calculus which provided the first clue on the existence
of vacuum nodes.  This interference suggests that perhaps
 the dilute gas approximation is  not inappropriate
for the $\theta=\pi$ regime.

Similar phenomena occurs in the O(3) $\sigma$--model which is
exactly solvable \cite{zam}. In a discrete formulation it is
equivalent by Haldane transformation \cite{Hal}
 to a spin $\half$ chain and for those chains the
Lieb-Schulz-Mattis theorem \cite{lms} establishes that there
are only two possibilities: either parity is spontaneously broken
or the mass gap vanishes. Since it is known that the mass gap is
zero for $\theta=\pi$ \cite{zamm}\cite{sh}, then parity is not
spontaneously broken in this model at $\theta=\pi$.

  In summary the analysis of CP symmetry contributes to shed
some light on the non-perturbative structure of  QCD at low
energies, and due to the  lack of exact results in QCD  this information
 becomes highly valuable. However, the nature and existence of a transition at
$\theta=\pi$ remains unveiled, although  the behavior
of the theory seems to be analogous to that of the
O(3) $\sigma$--model
\cite{new}. Numerical investigations of such a problem would be very
interesting to get new insights on the structure of  QCD vacuum.

 We thank  K. Gaw\c{e}dzki, H. Markum,  W. Sakuler  and G. Sierra
 for  discussions. This work was partially supported by the Human 
 Capital and Mobility Program of the U.E. (Grant No. ERBCHRX-CT92-0035) 
 and CICyT (Grant No. AEN94-0218).


\end{document}